\pdfoutput=1
\documentclass[10pt,a4paper]{article}
\usepackage{fix-cm}
\usepackage[T1]{fontenc}
\usepackage[utf8]{inputenc}
\usepackage[scaled=0.96]{uarial}

\pdfmapfile{+ua1.map}
\usepackage[a4paper,left=23mm,right=23mm,top=21mm,bottom=22mm,headheight=12pt,headsep=7mm,footskip=11mm]{geometry}
\usepackage{microtype}
\usepackage{xcolor}
\usepackage[largestsep]{titlesec}
\usepackage{fancyhdr}
\usepackage{multicol}
\usepackage{needspace}
\usepackage{xurl}
\usepackage{hyperref}
\definecolor{ink}{HTML}{20272D}
\definecolor{muted}{HTML}{5A636B}
\definecolor{titlered}{HTML}{8B2836}
\definecolor{rulegray}{HTML}{ABB2B8}
\definecolor{abstractgray}{HTML}{F7F8F8}
\hypersetup{colorlinks=true,linkcolor=ink,urlcolor=ink,citecolor=ink,
 pdftitle={The Human Guarantee: An open letter from emerging geoscientists on adapting to the generative AI revolution},
 pdfauthor={Davide Zaccagnino and co-authors},
 pdfsubject={Human oversight and responsibility in AI-assisted geoscience},
 pdfkeywords={artificial intelligence, geosciences, scientific responsibility, early-career researchers},
 bookmarksnumbered=false, pdfdisplaydoctitle=true}
\fancypagestyle{opening}{\fancyhf{}\fancyfoot[C]{\fontsize{12}{14}\selectfont\color{ink}\thepage}}
\titleformat{\section}{\fontsize{11.2}{14}\selectfont\bfseries\color{ink}}{}{0pt}{}
\titlespacing*{\section}{0pt}{16pt plus 3pt minus 2pt}{7pt}
\titleformat{\subsection}{\fontsize{10.5}{13.5}\selectfont\bfseries\color{ink}}{}{0pt}{}
\titlespacing*{\subsection}{0pt}{11pt plus 2pt minus 1pt}{5pt}
\newcommand{\affiliation}[2]{\par\noindent\hangindent=1.6em\hangafter=1\hypertarget{aff:#1}{}\makebox[1.6em][l]{\textsuperscript{#1}}#2\par}
\makeatletter
\renewcommand\@biblabel[1]{}
\renewenvironment{thebibliography}[1]
 {\section*{References}\fontsize{9}{12}\selectfont
  \list{}{\usecounter{enumiv}\setlength{\labelwidth}{0pt}\setlength{\labelsep}{0pt}
          \setlength{\leftmargin}{1.4em}\setlength{\itemindent}{-1.4em}
          \setlength{\itemsep}{5pt}\setlength{\parsep}{0pt}}
  \sloppy\clubpenalty10000\widowpenalty10000}
 {\endlist}
\makeatother
\begin{document}
\fontsize{10.5}{14.1}\selectfont
\color{ink}
\thispagestyle{opening}
{\fontsize{26}{30}\selectfont\bfseries\color{titlered}The Human Guarantee\par}
\vspace{4pt}
{\fontsize{13}{17}\selectfont An open letter from emerging geoscientists on adapting to the generative AI revolution\par}
\vspace{12pt}
\begingroup
\fontsize{9.2}{12.8}\selectfont
\setlength{\parskip}{0pt}
\raggedright
\mbox{Davide Zaccagnino\textsuperscript{\hyperlink{aff:1}{1},\hyperlink{aff:2}{2}}}, \mbox{Himanshu Agrawal\textsuperscript{\hyperlink{aff:3}{3}}}, \mbox{Malik Adam Alddoum Adam\textsuperscript{\hyperlink{aff:4}{4}}}, \mbox{José Antonio Bayona\textsuperscript{\hyperlink{aff:5}{5}}}, \mbox{Eric Beaucé\textsuperscript{\hyperlink{aff:6}{6},\hyperlink{aff:7}{7}}}, \mbox{Emanuele Biondini\textsuperscript{\hyperlink{aff:8}{8}}}, \mbox{Pavlos Bonatis\textsuperscript{\hyperlink{aff:9}{9}}}, \mbox{Emanuele Bozzi\textsuperscript{\hyperlink{aff:8}{8}}}, \mbox{Gaspard Farge\textsuperscript{\hyperlink{aff:10}{10}}}, \mbox{Rossella Fonzetti\textsuperscript{\hyperlink{aff:2}{2}}}, \mbox{Subash Ghimire\textsuperscript{\hyperlink{aff:11}{11}}}, \mbox{Octavi Gómez-Novell\textsuperscript{\hyperlink{aff:12}{12}}}, \mbox{Yihe Huang\textsuperscript{\hyperlink{aff:13}{13}}}, \mbox{Antonio Giovanni Iaccarino\textsuperscript{\hyperlink{aff:2}{2}}}, \mbox{Francesco Immorlano\textsuperscript{\hyperlink{aff:14}{14}}}, \mbox{Sara Klaasen\textsuperscript{\hyperlink{aff:15}{15}}}, \mbox{Folarin Kolawole\textsuperscript{\hyperlink{aff:16}{16}}}, \mbox{Linxuan Li\textsuperscript{\hyperlink{aff:17}{17}}}, \mbox{Zhonglan Liu\textsuperscript{\hyperlink{aff:18}{18}}}, \mbox{Leila Mizrahi\textsuperscript{\hyperlink{aff:19}{19}}}, \mbox{Angelica Monserrat Buenrostro\textsuperscript{\hyperlink{aff:20}{20},\hyperlink{aff:21}{21}}}, \mbox{Louisa Murray-Bergquist\textsuperscript{\hyperlink{aff:22}{22}}}, \mbox{Kadek Hendrawan Palgunadi\textsuperscript{\hyperlink{aff:23}{23}}}, \mbox{Marina Poposka Andreeva\textsuperscript{\hyperlink{aff:24}{24}}}, \mbox{Juan Porras\textsuperscript{\hyperlink{aff:25}{25}}}, \mbox{Giacomo Pozzi\textsuperscript{\hyperlink{aff:26}{26},\hyperlink{aff:2}{2}}}, \mbox{Henrique Berger Roisenberg\textsuperscript{\hyperlink{aff:27}{27}}}, \mbox{Pierre Romanet\textsuperscript{\hyperlink{aff:28}{28},\hyperlink{aff:29}{29}}}, \mbox{Boris Rösler\textsuperscript{\hyperlink{aff:30}{30},\hyperlink{aff:31}{31}}}, \mbox{Ilaria Spassiani\textsuperscript{\hyperlink{aff:2}{2}}}, \mbox{Jacek Szczygieł\textsuperscript{\hyperlink{aff:9}{9}}}, \mbox{Libin Tang\textsuperscript{\hyperlink{aff:32}{32},\hyperlink{aff:33}{33}}}, \mbox{Stefania Tarantino\textsuperscript{\hyperlink{aff:34}{34}}}, \mbox{Nikolaos Vavlas\textsuperscript{\hyperlink{aff:35}{35}}}, \mbox{Patricio Venegas-Aravena\textsuperscript{\hyperlink{aff:36}{36}}}, \mbox{Saiprasad Villiyanallur Raja\textsuperscript{\hyperlink{aff:1}{1}}}
\par\endgroup
\vspace{7pt}
{\fontsize{8.7}{11.5}\selectfont\color{muted}Affiliations on page~\pageref{affiliations}.\par}
{\fontsize{9.2}{12}\selectfont\textbf{Correspondence:} Davide Zaccagnino\enspace\href{mailto:davidezaccagnino@gmail.com}{davidezaccagnino@gmail.com}\par}
\vspace{12pt}
\begingroup
\setlength{\fboxsep}{11pt}
\setlength{\fboxrule}{0.45pt}
\noindent\fcolorbox{rulegray}{abstractgray}{%
\begin{minipage}{\dimexpr\textwidth-2\fboxsep-2\fboxrule\relax}
\fontsize{10}{13.5}\selectfont
\textbf{Abstract}\par\vspace{4pt}
Generative artificial intelligence (AI) can facilitate scientific research and accelerate the development of applications that benefit society. However, if institutional safeguards and incentives fail to adapt, faster scientific production can outpace our ability to understand and verify it. As emerging researchers in the geosciences, we call for a Human Guarantee: meaningful and critical human oversight, coupled with accountability for scientific claims, regardless of whether AI assisted their production. We advocate for the implementation of safeguards, alongside incentives that reward deep understanding, verification, and innovation to foster a healthy research environment where future scientists can thrive and public trust in science is preserved and strengthened.
\end{minipage}}
\endgroup
\vspace{5pt}
\section*{Who signs science?}

Artificial intelligence (AI) offers important opportunities to explore observations, analyse unprecedented amounts of data, accelerate and guide experiments, and advance our understanding of nature. \textbf{Generative AI can make conceptualisation, design, coding, and scientific writing more accessible}, helping emerging researchers learn and participate in scientific discovery. Nonetheless, generative AI, without oversight, can produce scientifically plausible content much faster than researchers can validate it, as the problem of fabricated references illustrates (Walters and Wilder, 2023). These issues predate generative AI, but the current pace of AI-driven science production can exacerbate them when institutional responses lag behind changes in research practice. Academic institutions, funders and publishers must adapt while managing competing commercial, academic, and public responsibilities. Our aim is to encourage the scientific community to use these tools responsibly while adapting training, evaluation and verification to the pace of change.

Science is made by and for people. As scientists, we hold our mandate on loan from society to investigate how natural systems work and transform our insights into benefits for society. Our credibility depends on the strength of the \textbf{human verification chain}, starting with authors and extending to reviewers, readers, and the broad range of stakeholders who trust what they read or apply (Thorp, 2023). This responsibility includes verifying consequential AI outputs against evidence and communicating their uncertainties. When research becomes an end in itself - counted, ranked and rewarded - the people it should serve recede from view.

Our concern is the \textbf{separation of authorship from responsibility}: poorly verified publication content may propagate into future studies, potentially entering future AI training datasets, undermining the reliability of subsequent research and increasing pressure on an already overworked peer review system (Drozdz and Ladomery, 2024). OpenAI's Navier-Stokes announcement and the subsequent preprint on AI autoformalisation illustrate the importance of checking that an automatically formalised and verified representation faithfully reflects the scientific claim being evaluated (OpenAI, 2026; Bastounis et al., 2026). This verification question is separate from the allocation of scientific credit.

Our response to these challenges must be timely and coordinated so that scientists can benefit from AI without losing the conditions needed to develop their expertise.

\section*{The times of science}

Scientific activity embraces different, overlapping realities, whose timescales shape how AI can support research and training. Textbook Science represents the core of tested knowledge and provides the backbone for training new generations of scientists. Applied Scientific Research uses sufficiently robust knowledge to address well-defined problems. Frontier Research investigates questions for which even the concepts and explanations may remain unsettled, and developing the right perspective and framework represents a great part of the effort.

Researchers move between these activities, whose aims, modalities, and timescales differ. At the frontier, research requires developing mental constructs, testing them against observations, and revising them through discussion. Reality exists beyond these descriptions, and evidence constrains them on a never-ending road to deeper understanding. This fundamental activity requires creative thinking, investigation, scientific discussion, lifelong training, and, most importantly, time. Relevant criteria include deeper understanding, innovation, the ability to support conclusions with compelling arguments, real-world data, and scientific vision. Applied research also generates new understanding, pursues public benefit, and connects academia, industry, and society. This field requires advanced tools and practical skills to efficiently and reliably explore new possibilities emerging from frontier research. Here, the metrics are reliability, speed, practicality, competitiveness, and usability. These priorities overlap; in both of these research activities, \textbf{AI-assisted findings require verification appropriate to the claims being made}, and neither activity has exclusive ownership of innovation or public benefit.

Generative AI represents a powerful tool for extending applied research horizons and accelerating production. Frontier research can also greatly benefit, as large language models (LLMs) have, for some tasks, substantial information-processing and pattern-recognition capabilities that \textbf{complement human expertise}. We must navigate this transition by fully employing the new technologies while preserving the conditions for emerging researchers to develop, establish and improve their skills.

\section*{Opportunities and challenges in geosciences}

Generative AI and conventional machine-learning and deep-learning tools have different roles in our field. Conventional methods can analyse large data volumes and interpolate within the conditions represented in their training data, while generative AI can support productivity in tasks such as coding and writing. Applications of machine learning range from analysing Earth observations for disaster management (Albayrak et al., 2026; Mulia, 2025; Juanara and Lam, 2025) to estimating volcanic eruption mass (Mousavi et al., 2026). We identify the following non-exhaustive opportunities and potential failure modes, without implying that all users or models encounter them. Several challenges predate AI; what changes is the scale of automation and the ease with which plausible outputs can be produced.

\textbf{1. AI can help identify and untangle relationships in the highly complex, hierarchical and interconnected Earth system, including patterns previously overlooked by humans.} Observed phenomena can often be explained by various physical mechanisms at different scales. However, statistical correlations do not establish physical causal relationships. This is a longstanding limitation of scientific inference. Generative AI adds the ability to connect diverse observations rapidly into seemingly plausible and self-consistent scientific narratives that lack a reliable, tested physical basis (Messeri and Crockett, 2024). The ease of generating such narratives increases the burden of checking AI-derived findings against observations and physical reasoning.

\textbf{2. AI is now widely used for the automated detection of weak signals (e.g., seismic phases, signals of slope instability) and their systematic monitoring.} However, as these methods become increasingly prevalent, researchers may come to view them as "black-box" tools - without adequately examining their assumptions, limitations, and potential biases. Such biases can propagate into subsequent analyses, influencing the reliability and replicability of scientific conclusions and hindering the ability of new researchers to build upon previous work. For instance, assuming that a deep-learning model for seismic phase picking ``just works'' can obscure missed arrivals and false picks, particularly outside its training conditions (Münchmeyer et al., 2022). Assessing its reliability requires additional checks, such as phase association checks that evaluate consistency across stations, comparison with manually reviewed earthquake catalogues, and manual inspection of both detected picks and waveforms where no picks were made. 

\textbf{3. AI models must contend with geological data that are often sparse, incomplete, or unevenly sampled, especially in deep-Earth studies, research in remote oceanic regions, and greenfield exploration (searching for minerals in unexplored areas) }(Schaaf and Bond, 2019; Ringler et al., 2022; Davies et al., 2025). Historical drill logs and field notes also reflect differences in interpretation and terminology among geologists (Hill et al., 2021). Models that interpolate successfully within their training data may still fail when extrapolating to poorly sampled geological settings. AI models can struggle with physical ambiguity and structural complexity when they lack contextual observations and domain knowledge supplied by fieldwork (Vafeas, 2026). These limitations are not new, but automated processing can reproduce them across many estimates unless data coverage and uncertainty are assessed explicitly (Rau et al., 2025). This is especially relevant to rare, potentially high-consequence geohazards such as large earthquakes or extreme floods, which may be underrepresented in training data.

\textbf{4. Using generative AI to explain poorly constrained cases (e.g., interpreting offshore geology from limited observations) can produce results strongly influenced by assumptions embedded in the prompt.} \textbf{These narratives are hypotheses, not additional observations.} Variations in prompt wording can alter model outputs and their apparent confidence, but this variability is distinct from uncertainty arising from incomplete geological evidence (Zhou et al., 2023). Checks against alternative prompts and independent evidence are therefore required.

\textbf{5. The growing popularity of AI encourages the development of new geoscientific applications, whose complexity should be justified by improvements over simpler approaches.} For instance, complex AI models applied to sparse and noisy datasets, which are frequent in geosciences, can lead to overfitting and ultimately induce biased interpretations where simpler approaches may be more robust. When methods perform comparably, the principle of parsimony (Occam's razor) favours the simpler approach. Methods should be evaluated against the scientific question, available data and suitable baselines (e.g., Zhuang, 2026), without discouraging exploratory uses that are clearly identified and independently tested (e.g., Schorlemmer et al., 2018).

\textbf{6. AI forecasting performance depends on the availability and representativeness of past records, as well as physical understanding.} AI has demonstrated strong performance in medium-range weather forecasting (Price et al., 2025), whereas reliable earthquake forecasting remains an unresolved challenge, particularly for rare, large events with limited historical records (Al Banna et al., 2020). Performance must therefore be assessed for each phenomenon, timescale and use. Reliance on historical data can limit forecasts under conditions differing from the past, including some future climate states. Event-specific evaluation can reveal limitations obscured by aggregate scores (Charlton-Perez et al., 2024), while tests across regions can expose limits to model transfer (Ghimire and Guéguen, 2024). Incorporating physical constraints, evaluating performance against out-of-sample data, and quantifying uncertainty can improve reliability.

\textbf{7. AI-based classification methods, including deep-learning models, require careful assessment of uncertainty in the resulting geoscientific estimates.} The probability assigned to a class does not necessarily quantify uncertainty in the final scientific estimate; for example, a confident seismic phase pick does not guarantee a well-constrained earthquake location. Pick errors, network geometry and velocity-model uncertainty must also be considered. Class probabilities should be calibrated against observations, and uncertainties from measurements and model assumptions propagated through subsequent calculations.

\textbf{8. AI can support the development of data-analysis code, especially during the initial testing of a scientific hypothesis, and promote the reproducibility and replicability of these analyses.} AI can help develop computer programs efficiently to explore scientific hypotheses. Moreover, AI significantly facilitates the systematic development of tests of code logic as well as benchmarking against existing models and well-known problems. \textbf{AI users must verify and understand AI-generated code.} Distributing this code as well-documented, well-structured and user-friendly open-source packages promotes peer review of the source code and the reproduction and replication of the analyses or simulations supporting research. Tools such as floatCSEP illustrate the value of repeatable earthquake-forecast testing (Iturrieta et al., 2026). AI can help revisit and check existing codes, improving their usability and accessibility to a broader community, and even combine them to create new workflows. Thus, AI can promote Findable, Accessible, Interoperable, and Reusable (FAIR; Wilkinson et al., 2016) scientific practices. As in the pre-AI era, successful computer program execution alone does not establish scientific validity.

\section*{What we risk losing}

\subsection*{1. The ability and willingness to think deeply}

Quick and rewarding answers can make us accustomed to obtaining solutions without applying much effort ourselves. When AI replaces independent intellectual effort rather than supporting it, some users may learn and remember less (Bastani et al., 2025; Barcaui, 2025), although findings from specific educational studies and workplace surveys should not be generalised to all geoscientists (Lee et al., 2025). Delegating tasks to tools is nothing new: computers and other technological tools can also support learning (Risko and Gilbert, 2016; Ellington, 2003). The concern is uncritical reliance: improved misinformation judgements during AI-assisted sessions did not translate into lasting independent discernment in one study (Rani et al., 2026). Active reasoning and training can help \textbf{balance assistance with learning} (Kabashkin, 2025), and carefully designed tutoring safeguards can mitigate adverse effects (Bastani et al., 2025).

\subsection*{2. The substance of scientific dialogue}

The interactive interface of an LLM creates the impression of a dialogue, but an exchange that accommodates our assumptions may give us confidence before we have worked through the problem ourselves (Messeri and Crockett, 2024). Related effects have been documented for Internet searches (Fisher et al., 2015). LLMs draw on patterns learnt during training and the context we provide, adjusting their responses accordingly. Colleagues bring experience, convictions, and questions that we cannot reshape by changing our instructions. This friction requires us to explain ourselves, reconsider our arguments, and sometimes abandon them. AI can help us prepare for and extend these discussions, but the risk arises when \textbf{accommodating responses substitute for independent criticism}.

\subsection*{3. Time to learn, innovate and enjoy discovery}

Giambattista Vico linked knowing with making (Vico, 1988). When careers depend on publication rate and visibility, productivity becomes both the metric and the goal. Emerging researchers may feel compelled to use AI to keep pace, even when they need more time to develop and understand their results. This institutional pressure existed before generative AI, but faster production can intensify it, leaving less room for learning, exploration, and the pleasure of discovery. If output expectations absorb the time saved by AI, researchers may also have less \textbf{time to study existing work, question assumptions and develop new ideas}. Working through a difficult problem can build confidence and scientific judgement; AI can support this process if improved efficiency protects time for inquiry, experiments and fieldwork. Gross et al. (2026) report that US graduates in science, technology, engineering and mathematics (STEM) whose dissertations contained detected AI-generated writing were less likely to enter academic research careers. This observational association does not establish that AI use caused those career outcomes, but it suggests that academically oriented students may still perceive AI as a competitor in solving their own scientific questions and not as a tool to explore them more effectively, deeply and rapidly. 

\subsection*{4. The continuity of scientific training}

Scientific expertise develops through years of practice, mistakes, and discussion with more experienced colleagues (Ma et al., 2020). If institutions use AI mainly to replace work usually done by emerging researchers, opportunities for supervised practice could shrink, especially where established scientists control funding, research directions and professional networks. Emerging researchers could then lose opportunities to work, learn and develop professional independence. Our concern is that researchers may produce scientific documents without the knowledge needed to question and defend their contents. This is not inevitable: automating routine work can instead create \textbf{time for mentoring and more demanding scientific tasks}. Another potential risk is that if researchers benefit from using AI without sufficient care, for example, to produce more publications, others may find themselves at a career disadvantage or feel compelled to join the same arms race.

\subsection*{5. Access to frontier research and neutrality of scientific development}

Access to instruments, computing facilities, and technical support has long shaped research opportunities. Paid AI subscriptions and computing requirements can create another barrier between institutions that can afford these resources and those that cannot, thereby increasing \textbf{funding inequality}. The companies that provide proprietary models also determine their prices, access conditions, and training methods (Ahmed et al., 2023). This can increase dependence on commercial infrastructures whose priorities generally do not align with those of the researchers (Ahmed et al., 2023; Varoufakis, 2023). Risks include reduced scientific independence and reliance on potentially biased commercial models, motivating institutional access agreements and shared alternatives.

\subsection*{6. Responsibility for what we sign}

Signing a paper means accepting responsibility for the integrity of its methods, evidence, and conclusions. This connection is weakened when authors submit AI-generated claims without adequately understanding or verifying them, giving credibility to work they may be unable to defend (Thorp, 2023). As Hans Jonas argued in his major work, The Imperative of Responsibility: In Search of an Ethics for the Technological Age (Jonas, 1984), it is the magnitude of power that we have arrogated to ourselves that imposes the obligation of acting responsibly upon us. AI can accelerate content generation without a corresponding increase in verification capacity. \textbf{Accountability remains with the authors}, who must use defensible procedures and communicate uncertainty honestly; authorship cannot guarantee that every conclusion will remain correct.

\subsection*{7. The reliability of scientific literature}

The dissemination of unverified results has the potential to compromise the integrity of future research. Once published and repeatedly cited, an initial error may become more difficult to identify and correct; \textbf{citation does not constitute replication}. AI can accelerate this propagation if unreliable material is reused in literature synthesis or model training. Separately, Shumailov et al. (2024) demonstrated that repeated training on generated data can degrade model performance under particular conditions. A further risk arises if organisations with an economic interest in a particular conclusion employ AI to produce substantial amounts of misleading material. The act of repetition could create the impression of agreement while meticulous rebuttals require a considerable investment of time. Historical efforts to manipulate scientific communication for commercial gain underscore the necessity of explicitly recognising this possibility (e.g., Supran et al., 2023).

\subsection*{8. The integrity of scientific peer review}

The process of accurate peer review is laborious and time-consuming. As submissions become easier to produce, the effort required for verification may exceed the time available to reviewers. The use of AI to write reviews (Liang et al., 2024) or screen submissions (Springer Nature, 2025) does not eliminate the necessity for individuals to understand and take responsibility for the assessments conducted. Reviewers assess evidence and argument within the limits of their expertise and the material available; they cannot guarantee correctness. The practical safeguards proposed below include \textbf{transparent reports, adequate review time and resources, and clear rules for permitted AI use}.

\subsection*{9. The public trust that makes science possible}

Our research, especially in hazard science, can directly affect people's lives, for instance, by influencing how civil protection prepares and which environmental risks are considered acceptable. An unverified estimate can therefore have consequences far beyond the research output in which it appears. Our responsibility begins with the observations on which our claims rest. Reliable science requires instruments that are maintained, people who know how to use them, and data whose origin, processing and uncertainty are documented (Wilkinson et al., 2016; UNESCO, 2021). The technicians, engineers and scientists who sustain this work must receive support. Public trust also depends on \textbf{transparent communication of uncertainty, acknowledgement of errors and respectful engagement with affected communities}, as recognised in the Cape Town Statement on Geoethics (Di Capua et al., 2017). AI-assisted communication must meet these same responsibilities.

\subsection*{10. Cross-disciplinary collaboration}

LLMs can produce apparently professional code from relatively simple prompts. A geoscientist may create seemingly high-quality cloud services and software, which would require knowledge of computer science (CS) concepts such as data engineering, databases and systems integration. However, the geoscientist may lack the expertise needed to evaluate the generated logic adequately. Under the pressure of funding that demands tangible deliverables, geoscientists may be tempted to use AI models for this purpose with limited cooperation with CS experts. Similarly, geoscientists may use generative AI to implement complex models reaching beyond their expertise without seeking support from, or interacting with, appropriate experts (whether they be other geoscientists, physicists or mathematicians). Delegating cross-disciplinary tasks to a tool may be less costly than employing other scientists, but may weaken cross-disciplinary collaboration if expert review is bypassed. Shared AI tools can also facilitate collaboration, provided \textbf{relevant specialists help frame the problem and validate the solution}.

\section*{Recommendations for action}

The most urgent need is to make existing authorship responsibilities effective under faster AI-assisted production: every author must accept responsibility for their contribution and help resolve questions about the work's integrity through transparent disclosure, appropriate verification, and communication of uncertainties. \textbf{Scientific authorship requires intellectual responsibility.}

Safeguards are necessary when production overwhelms verification; however, rules can work only if \textbf{incentives make responsible behaviour viable and rewarding}. We propose the following practical steps for researchers, institutions, funders and journals:

\subsection*{1. Strengthen author accountability and research transparency}

Before accepting a submission, journals and platforms should request \textbf{clear author-contribution and AI-use statements}, together with confirmation that all authors approved the final manuscript. For scientifically consequential uses, AI-use statements should identify the tools and versions used, their role, and how outputs were checked; routine language editing may warrant simpler disclosure, consistent with emerging geoscience guidance (Lacassin et al., 2026). Authenticated ORCID identifiers can support attribution where available, but neither an identifier nor an academic lineage establishes scientific validity (ORCID, n.d.). Editorial scrutiny should be proportionate to methodological risk, with clear criteria and the right to appeal.

\subsection*{2. A mindset change towards publishing}

The FAIR Guiding Principles encourage researchers to make digital outputs - including data, software code, algorithms, and metadata - Findable, Accessible, Interoperable and Reusable (Wilkinson et al., 2016). Journals should ask authors to provide documented data, code, source files and validation tests, where applicable. For AI-assisted research, relevant model versions, settings, prompts and processing steps should support independent evaluation of the reported results. Computational reproducibility alone does not establish the validity of scientific assumptions or interpretations. For restricted data, secure or controlled-access verification procedures should be available when legally and ethically permissible; FAIR does not require unrestricted access. This work requires time and support for authors and reviewers. Institutions should therefore reward its quality, building on existing research-assessment reforms (DORA, 2013; Arentoft et al., 2022). \textbf{Fewer, higher-quality, better-supported publications} can benefit the scientific community and provide an improved corpus for future LLMs.

\subsection*{3. Promote inclusive access to generative AI}

Generative AI can help reduce inequalities in scientific communication arising from documented language barriers (Amano et al., 2023; Testani et al., 2025), particularly for researchers whose first language is not English. Tools for translation, grammar correction and text editing can help bridge existing gaps for researchers from underrepresented scientific regions, making scientific communication more inclusive and accessible. Institutions should therefore promote \textbf{equitable access to these technologies and training} in their use while ensuring that their use supports, rather than replaces, researchers' scientific contributions. Authors must check that editing preserves scientific meaning.

\subsection*{4. Reward peer review}

As AI-assisted production increases demands on verification, reviewing should receive \textbf{allocated workload time}, appropriate monetary or non-monetary compensation, and career recognition, regardless of whether the recommendation is positive or negative. Journals should verify reviewers' expertise and independence to support fair assessment. Journals should require clear conflict-of-interest declarations. Permitted AI assistance should be disclosed, and confidential manuscripts should be protected in accordance with journal policy. Editors remain responsible for publication decisions.

\subsection*{5. Simplify journal and project submission procedures}

We propose using AI assistance for administrative, consistency, and reporting checks at pre-screening and post-revision stages, with \textbf{qualified humans controlling consequential scientific decisions}. Journals and funders could accept flexible initial formats and offer optional assistance with reference checks, figure labels and conversion to their required style, with authors approving changes. This would reduce repetitive work without imposing one universal template or restricting scientific expression. A common minimum reporting standard should document AI use, verification procedures, data and code availability, and responsibility for the final conclusions. Confidential material should be processed only under approved data-protection arrangements.

\section*{A long-term vision}

\subsection*{6. Put scientific understanding at the core of assessment}

Oral discussion should complement written assessment, from school examinations to research degrees and academic appointments. Candidates should demonstrate their ability to pose insightful questions and explain, challenge and extend their work, including by performing appropriate tasks without AI assistance. As AI makes polished outputs easier to produce, assessment should examine candidates' knowledge, their ability to apply it, the innovation of their research and their vision (DORA, 2013; Arentoft et al., 2022). It should also examine how candidates check AI outputs, with accessible alternatives where oral examination is unsuitable. Careers in frontier research should depend on \textbf{demonstrated understanding and the quality of contributions}; publication counts and citation metrics should inform, rather than completely drive, this assessment.

\subsection*{7. Reward teaching and mentoring}

Becoming scientists requires practice, discussion and learning from mistakes and failures. We should safeguard intergenerational collaboration and address the vulnerabilities of early-career researchers by introducing incentives for \textbf{high-quality mentoring and teaching} (Ma et al., 2020; Arentoft et al., 2022). AI-assisted work should include opportunities to explain methods, examine errors and develop independent judgement. Universities need funding for observations, laboratories and scientific training; attracting funds should support these purposes without becoming the main measure of academic value.

\subsection*{8. Fund and protect frontier research}

Applied and frontier research both need resources proportionate to their goals, including competitive grants for ambitious projects. Frontier research requires \textbf{stable public and private support to protect long-term inquiry}, observations and training from repeated interruptions and conflicts of interest. As AI changes research workflows, shared infrastructures and equitable access can reduce dependence on individual providers (Public AI Network, 2024; Current AI, n.d.), alongside training and stronger protection of users' data.

\subsection*{9. Rebuild a culture of scientific discussion}

Scientific ideas need time to be questioned, defended and reconsidered. At some large conferences, presentations last a few minutes with little room for questions, and discussion becomes something to fit into the remaining time.

Some of the \textbf{time saved through AI-assisted routine work could be reserved for scientific discussion}, provided institutions value this exchange alongside publication output. We therefore encourage meetings, both onsite and online, on a human scale, organised for scientists and focused on the exchange of ideas, with enough time to examine our claims through dialogue (Plato, Theaetetus, 150b--d). AI can help participants prepare and follow up discussions, while colleagues provide independent criticism and new perspectives.

\subsection*{10. Communicate the societal value of scientific expertise}

As frontier researchers, our societal duty includes producing and validating knowledge, developing applications and explaining their implications. A central human contribution is the ability to interpret, critically evaluate, and take responsibility for geoscientific claims. Indeed, the role of geosciences and their impact on society will continue to matter even if AI systems perform an increasing share of scientific tasks. The advent of generative AI may allow us to devote more time to studying and exploring the boundaries of knowledge, provided institutions support learning and independent inquiry. In the future, it will be key to communicate the need for \textbf{a strong community of scientists who understand how the Earth works}.

\section*{Conclusion}

The current era equips us with instruments that can greatly extend scientific exploration. The concern is that, while we seek rapid solutions, we lose sight of why knowledge matters, what it is for, and who is responsible for it. Publishing AI-assisted claims without adequate verification obscures responsibility: tools can assist with scientific work, but their use does not remove authors' duty to evaluate and defend the results. We call on researchers, institutions, funders and journals to protect science by making verification and content responsibility effective conditions for authorship. \textbf{Institutions must protect the time and training needed to meet these responsibilities.} By adapting together, we can use AI's capabilities to \textbf{strengthen scientific understanding and public trust.} That is the Human Guarantee.

\vspace{4pt}
\section*{Acknowledgements}

The authors thank Carlo Doglioni, Francesco Grigoli and Didier Sornette for stimulating discussions.

\clearpage
\section*{Author affiliations}
\phantomsection\label{affiliations}
\begingroup
\fontsize{8.6}{11.4}\selectfont
\setlength{\parskip}{5pt plus 1pt}
\begin{multicols}{2}
\raggedright
\affiliation{1}{Institute of Risk Analysis, Prediction and Management (Risks-X), Academy for Advanced Interdisciplinary Studies, Southern University of Science and Technology (SUSTech), Shenzhen, Guangdong, China}
\affiliation{2}{National Institute of Geophysics and Volcanology (INGV), Rome, Italy}
\affiliation{3}{School of GeoSciences, The University of Edinburgh, Edinburgh, UK}
\affiliation{4}{Department of Mathematics and Computer Sciences, Physical Sciences and Earth Sciences, University of Messina, Messina, Italy}
\affiliation{5}{School of Earth Sciences, University of Bristol, UK}
\affiliation{6}{Lamont-Doherty Earth Observatory, Columbia University, USA}
\affiliation{7}{Institut des Sciences de la Terre, Université de Grenoble Alpes, France}
\affiliation{8}{Department of Earth Sciences, University of Pisa, Pisa, Italy}
\affiliation{9}{Institute of Earth Sciences, University of Silesia, Sosnowiec, Poland}
\affiliation{10}{Université Paris Cité, Institut de physique du globe de Paris, CNRS, F-75005 Paris, France}
\affiliation{11}{IMSIA (UMR 9219), CNRS, EDF, ENSTA Paris, Institut Polytechnique de Paris, Palaiseau, France}
\affiliation{12}{Department of Earth and Ocean Dynamics, University of Barcelona, Barcelona, Spain}
\affiliation{13}{Department of Earth and Environmental Sciences, University of Michigan, USA}
\affiliation{14}{CMCC Foundation - Euro-Mediterranean Center on Climate Change, Italy}
\affiliation{15}{Research School of Earth Sciences, Australian National University, Australia}
\affiliation{16}{Department of Earth \& Environmental Sciences, Columbia University, USA}
\affiliation{17}{Seismological Laboratory, Division of Geological and Planetary Sciences, California Institute of Technology, Pasadena, CA, USA}
\affiliation{18}{College of Earth Sciences, State Key Laboratory of Deep Earth Exploration and Imaging, Jilin University, Changchun, China}
\affiliation{19}{Swiss Seismological Service, ETH Zurich, Zurich, Switzerland}
\affiliation{20}{Department of Structural and Geotechnical Engineering, Pontificia Universidad Católica de Chile, Santiago, Chile}
\affiliation{21}{Section 2.6 Seismic Hazard and Risk Dynamics, GFZ Helmholtz Centre for Geosciences, Potsdam, Germany}
\affiliation{22}{Department of Earth Science, Vrije Universiteit Amsterdam, Amsterdam, Netherlands}
\affiliation{23}{Department of Geophysical Engineering, Institut Teknologi Sepuluh Nopember, Surabaya, Indonesia}
\affiliation{24}{Institute of Earthquake Engineering and Engineering Seismology, Ss. Cyril and Methodius University in Skopje, North Macedonia}
\affiliation{25}{Department of Earth Sciences, University of Geneva, Geneva, Switzerland}
\affiliation{26}{Department of Geosciences, Padova University, Italy}
\affiliation{27}{Institut für Geowissenschaften, Christian-Albrechts-Universität zu Kiel, Kiel, Germany}
\affiliation{28}{Observatoire de la Côte d'Azur, Université Côte d'Azur, CNRS, IRD, Géoazur, Valbonne, France}
\affiliation{29}{Cerema, DRN, DTerMed, Valbonne, France}
\affiliation{30}{Center for Scientific Research and Higher Education at Ensenada, Mexico}
\affiliation{31}{University of the Andes, Colombia}
\affiliation{32}{Department of Civil Engineering, McMaster University, Hamilton, Canada}
\affiliation{33}{Department of Earth Sciences, Uppsala University, Uppsala, Sweden}
\affiliation{34}{National Institute of Geophysics and Volcanology (INGV), L'Aquila, Italy}
\affiliation{35}{Department of Geophysics, Aristotle University of Thessaloniki, Greece}
\affiliation{36}{Department of Engineering, Universidad Loyola Andalucía, Seville, Spain}
\end{multicols}
\endgroup
\end{document}